\documentclass{eas}
\usepackage{graphicx}
\newcommand{\lsun}{\mbox{$L_\odot$}}
\newcommand{\msun}{\mbox{$M_\odot$}}
\newcommand{\hii}{H\mbox{\sc ~ii} }
\TitreGlobal{To appear in: Astronomy in the submillimeter and far infrared domains
 with the Herschel Space Observatory}
\begin{document}
\title{Our knowledge of high-mass star formation \\
at the dawn of Herschel} 
\runningtitle{Motte \& Hennebelle: High-mass star formation at the dawn of \emph{Herschel}}
\author{Fr\'ed\'erique Motte}\address{Laboratoire AIM, CEA/DSM - CNRS - Universit\'e Paris Diderot, DAPNIA/SAp, CEA-Saclay, 91191 Gif-sur-Yvette Cedex, France}
\author{Patrick Hennebelle}\address{\'Ecole normale sup\'erieure, 24 rue Lhomond, 75005 Paris, France}
\begin{abstract}
We review the theories and observations of high-mass star formation emphasizing the differences with those of low-mass star formation. We hereafter describe the progress expected to be achieved with \emph{Herschel}, thanks notably to  Key Programmes dedicated to the earliest phases of high-mass star formation.   
\end{abstract}
\maketitle

\section{Introduction}

High-mass stars, also called OB stars, have luminosities larger than $10^3~\lsun$, spectral types B3 or earlier, and stellar masses roughly spanning the range $10-100~\msun$. From their birth to their death, high-mass stars are known to play a major role in the energy budget of galaxies via their radiation, their wind, and the supernovae. Despite that, the formation of high-mass stars remains an enigmatic process, far less understood than that of their low-mass (solar-type) counterparts. 

Theoretically, the copious UV flux emitted by a stellar embryo of more than $8~\msun$ heats and ionizes its parental molecular cloud, leading to the formation and development of a hot core and an \hii region. These physical and chemical feedback processes are difficult to treat but must be added in  analytical and numerical simulations classically used for low-mass star formation. These further difficulties, summing up with those associated with the star formation process itself (see chapter on low-mass star formation), have long delayed any effort in this direction.

Observationally, the main difficulty arises from the fact that high-mass stars are fewer in number than low-mass stars (see the shape of the IMF in, e.g., Fig.~2 of the chapter on low-mass star formation). Therefore, molecular clouds able to form high-mass stars are statistically more remote (typically at $d_{\mbox{\scriptsize Sun}}> 1$~kpc) than those of well-studied low-mass star-forming regions. Current observational studies of high-mass star formation therefore suffers from a lack of spatial resolution and from our bare knowledge of remote star-forming regions.

The past ten years have seen an increasing interest in approaching the issue of the formation of high-mass stars, from both the theoretical and observational sides. Here we review the recent progress made in this domain, and especially in the context of \emph{Herschel} programmes.

\section{Formation of high-mass stars: theory}

\subsection{Overview}

One of the main differences between the formation of  high-mass and low-mass stars is that the radiation field of a massive protostar plays a more important role. Indeed, the massive stellar embryo strongly heats the gas and even prevents further matter accretion through its radiation pressure. From a theoretical point of view, this  implies that the radiative transfer must be treated in parallel to the hydrodynamics. This represents a severe complication mainly responsible for the limited numbers of theoretical studies of this process.

Here is presented a short introduction to the theory of high-mass star formation. The reader interested in this topic will find more complete reviews in Zinnecker \& Yorke (\cite{ZY07}) and Beuther et al. (\cite{beut07}). In the following we exclusively focus on  studies specific to the formation of massive stars, keeping in mind that most of them are modified versions of theories developed for the formation of low-mass stars (see, e.g., the chapter on low-mass star formation). We first describe the basic principles used to estimate the largest stellar mass that one expects to form in the presence of  radiative forces. We then present the various numerical simulations and alternative scenarios which have been performed and proposed to explain the formation of high-mass stars.

\subsection{Basic principles and simple estimates}

The first estimates of the largest stellar mass that can possibly be assembled are due to Larson \& Starrfield (\cite{larson-starrfield}) and Kahn (\cite{kahn}). The principle of their analysis is to compare the radiative pressure of a massive stellar embryo to the ram pressure induced by the gravitational collapse of its surrounding massive cloud, in its inner and outer parts. If the luminosity of the central star becomes high enough, the radiation pressure may become important and prevent further accretion onto the central object. Since the radiation pressure is acting on the dust grains, one has to assume that the frictional coupling between the gas and the dust is sufficiently strong so that forces acting on the dust grains are transmitted to the gas. 

In the inner part of the collapsing cloud, the temperature becomes high and the dust grains evaporate. There is thus a dust shell whose inner edge is located at the radius, $r$, where the grains evaporate. At this radius, the radiation pressure is $L_\star / 4 \pi r^2 c$, where $L_\star$ is the stellar luminosity and $c$ the speed of light. The dynamical pressure is $\rho u^2$, where $\rho$ is the density and
$u$ the infall speed which is given by $u^2 \simeq 2 G M_\star / r$, where $G$ is the gravitational constant and $M_\star$ the mass of the protostar. This leads to the ratio of radiative to ram pressures:
	\begin{eqnarray}
	\Gamma  = { L_\star / 4 \pi r^2 c \over \rho u ^2}
	\simeq 1.3 \times 10^{-11}~ { L_\star / \lsun \over (M_\star / \msun) ^{1/2}} ~r^{1/2}. 
 	\end{eqnarray}
Using an analytic estimate for the temperature inside the cloud and based on the assumption that the grains evaporate at a temperature of $\sim$1\,500~K, Larson \& Starrfield (\cite{larson-starrfield}) estimate the radius of the shell to be:
	\begin{eqnarray}
	r \simeq 2.4 \times 10^{12} {(L_\star / \lsun) ^{1/2} \over (M_\star/\msun)^{1/5}} \, {\rm cm} 
	\simeq 3.3~ {(L_\star /10^3~\lsun) ^{1/2} \over (M_\star/8~\msun)^{1/5}} \, {\rm AU}.
	\end{eqnarray}
It follows from Eqs.~2.1--2.2 that 
	\begin{eqnarray}
	\Gamma \simeq 2 \times 10^{-5}~ {(L / \lsun) ^{6/5} \over (M/\msun)^{3/5}}.
	\end{eqnarray}
For a stellar mass of $20~\msun$, corresponding to a luminosity of about $4 \times 10^4~\lsun$, $\Gamma$ roughly equals unity. Therefore, according to Larson \& Starrfield (\cite{larson-starrfield}), the mass at which radiative pressure impedes accretion is around $20~\msun$. 

In the outer part of the collapsing cloud, the radiative pressure becomes equal to $\chi L / 4 \pi r^2 c$, where $\chi$ is the opacity due to dust grains and averaged over the stellar spectrum. The estimate done by Larson \& Starrfield (\cite{larson-starrfield}) is uncertain since the above expression entails the dust opacity and since the grains properties were poorly known by the time of their study. A more accurate estimate has been done by Wolfire \& Cassinelli (\cite{WC87}) by using the optical properties and composition of the mixture of dust grains proposed by Mathis et al. (\cite{mathis}). Assuming an accretion rate of $10^{-3}~\msun\,$yr$^{-1}$ in a $100~\msun$ cloud, Wolfire \& Cassinelli show that $\Gamma$ is larger than one for any reasonable value of the radiation temperature. They conclude that building a massive star with the ``standard'' dust grain mixture (cf. Mathis et al. \cite{mathis}) is difficult and requires reducing the grain abundance by large factors ($\sim$4--8). They thus propose, as a solution to the high-mass star formation problem, that the dust abundance could be locally decreased by an external shock or an internal ionization front.

\subsection{Radiative hydrodynamical simulations}

More recently, multidimensional numerical simulations have been performed, treating the radiation and the dynamics self-consistently. 
In these studies, it has been assumed that the radiation arises from both the accretion and the stellar luminosity. While the former is dominant during the earliest phase of the collapse, the latter rapidly becomes more important. Since computing the radiative transfer is in general a difficult problem, these calculations have made the diffusion approximation (e.g. Mihalas \& Mihalas \cite{mihalas}). We will see below that the frequency dependence of the dust opacity which was neglected in Sect.~2.2 is an important issue.  
 
One of the main motivations of these calculations is to determine whether the presence of a centrifugally supported optically thick disk, inside which the radiative pressure would be much reduced, may allow to circumvent the radiation pressure problem.

\subsubsection{Bidimensional multi-wavelengths calculations}

Bidimensional axisymmetric numerical simulations have been performed by Yorke \& Sonnhalter (\cite{YS02}) in the frequency dependent case (using 64 intervals of frequency) and in the gray case (one single interval of frequency). The cloud they consider is centrally peaked, has a  thermal over gravitational energy ratio of about 5$\%$ initially, and is slowly rotating. They have explored 3 different cases corresponding to 30, 60, and $120~\msun$ cores.

The fiducial run corresponds to a $60~\msun$ core treated in the frequency-dependent case. After $\sim$10$^5$~yr, the central core has a mass of about $13.4~\msun$ and the surrounding cloud remains nearly spherical. After $\sim$2$\times 10^5$~yr, the mass of the central core is about $28.4~\msun$ and the cloud starts to depart from the spherical symmetry. In particular, the infall is reversed by radiative forces in the polar region while the star continues to accrete material through the equator where the opacity is much higher. This effect which has first been proposed by Nakano et al. (\cite{nakano}) and Jijina \& Adams (\cite{jijina}) is  known as the ``flashlight effect''.  Once  the stellar mass has grown to about $33.6~\msun$, the central star is no longer accreting although $30~\msun$ of gas is still available within the computational grid. The infall is then reversed in every directions indicating that the radiative forces are effectively preventing further accretion. 

If instead of a multi-frequency treatment, the gray approximation is made, the early evolution is similar but becomes notably different after $\sim$2.5$\times 10^5$~yr. In particular, there is no evidence of any flow reversal. Instead the material flows along a thin disklike structure, supported in the radial direction by both centrifugal and radiative forces. At the end of the simulation, the mass of the central star is about $20.7~\msun$.

When the mass of the cloud is initially $30~\msun$, all the mass of the cloud is finally accreted onto the central star.  For a cloud mass of $120~\msun$, as soon as a stellar mass of $32.9~\msun$ has been accreted, reversed flows develop in the polar direction but the central object continues to accrete an additional $10~\msun$ via an equatorial flow through a disklike structure.

\subsubsection{Tridimensional calculations}

Very recently, tridimensional calculations have been performed by Krumholz et al. (\cite{krumholz}) using the gray approximation for the radiative transfer. Their initial conditions (aimed at reproducing the model of McKee \& Tan (\cite{mckee}) consist in a  centrally peaked $100~\msun$ cloud with a density profile proportional to $r^{-2}$. The initial turbulence within the cloud is sufficient to ensure an approximate hydrostatic equilibrium. Turbulent motions first delay the onset of collapse but, as the turbulence decays, the cloud starts to collapse. Comparison is made with runs for which a barotropic equation of state (EOS) is used (an isothermal EOS for most of the gas but an adiabatic EOS at high densities).

In particular, Krumholz et al. (\cite{krumholz}) find that, when the radiative transfer is taken into account, the gas temperature inside the cloud is higher than in the barotropic case, by factors up to 10, which are depending on the cloud density. Moreover, while the morphology of the flows in the radiative and barotropic cases are similar on large scales, they are significantly different on small scales. Because of the lack of thermal support, the gas in the isothermal run is much more filamentary, disks are flatter, filaments have smaller radii, and shock structures are thinner. As a consequence, the cloud is fragmenting much less when  radiation is taken into account than when a barotropic equation of state is used. In the radiative transfer case, only a few condensations are indeed able to survive, the others being sheared apart in the protostellar disk.

\subsection{Alternative scenarios} 

\subsubsection{Competitive accretion}

The above-mentioned calculations have been performed in the context of a monolithic collapse, that is to say considering the collapse of a single, initially centrally-peaked cloud. In the alternative scenario of competitive accretion, several stellar embryos are building up their mass within a common cloud also called protocluster (Bonnell et al. \cite{bonnell1}, \cite{bonnell3}). The accretion of the protostars depends on their location: it is strong at the cloud center where the gas density is high while it is weak near the cloud edge where less gas is available. The accretion also depends on the mass of the stars since the most massive stars are able to attract more gas and therefore increase further their accretion domain. The specificity of the competitive accretion compared to a monolithic collapse is that the gas is not assumed to be gathered before the star formation begins. In fact, competitive accretion provides the main physical mechanism to gather the matter. In the scenario of the competitive accretion, high-mass stars are forming in the center of the protoclusters. Indeed, the gravitational force of the protocluster inside which stars are competitively accreting attracts a significant fraction of the gas (and the stars) towards the cluster center. 

A recent observational and theoretical study of a $\sim$10$^3~\msun$ clump in NGC~2264 suggests an intermediate situation between the competitive accretion and the massive turbulent core model (Peretto et al. \cite{peretto1, peretto2}). The NGC~2264 clump seems indeed to be globally collapsing while well-defined condensations independently develop and collapse. Moreover, the gravitational merging of a few Class~0 protostellar cores may be occurring in the center of the massive clump.

\subsubsection{Stellar collisions and mergers}

In this scenario, originally proposed by Bonnell et al. (\cite{bonnell2}), the massive stars form by the coalescence of a large number of low-mass protostars.  The original motivations for proposing stellar collisions as a formation process for massive stars was, first, to circumvent the radiative pressure effect and, second, to explain the large number of massive stars in dense stellar clusters. Today, the first problem is known to be alleviate by the flashlight effect (see Sect.~2.3.1) whereas the second remains for a few exceptional cases of high-mass star formation. 

To estimate the star density required for the stellar collisions to be important, it is necessary to compare the collision time scale and the stellar evolution time. The former can be estimated using the stellar collision time per star (Binney \& Tremaine \cite{binney}). Typical stellar densities of about $10^7-10^8$~pc$^{-3}$ are required to set reasonable values of the mass, size, and velocity dispersion for the stellar collision partners. Non-equal mass encounters, circumstellar disks, and binary components could decrease this number to about 10$^6$ pc$^{-3}$, but such stellar density is highly unlikely to be the general case of high-mass star-forming sites.

\section{Formation of high-mass stars: observations}

\subsection{Open questions}

The fundamental unknown of high-mass star formation remains the physical process leading to stellar masses larger than $8~\msun$. As explained in Sect.~2, there are two competing families of models based on either accretion or coalescence with currently a slight advantage for the accretion scenarios. Observations can help end this debate by giving strong contraints on the evolutionary sequence and the initial conditions of OB star formation. We summarize below our limited knowledge of these topics and list related open questions (see also Beuther et al. \cite{beut07}).

\subsubsection*{What is the evolutionary sequence leading from molecular clouds to OB stars?}

Unlike for low-mass stars (see, e.g., Fig.~1 of the chapter on low-mass star formation), there is no observational evolutionary sequence which is firmly established for masssive young stellar objects (massive YSOs). Several attempts to derive evolutionary scenarios have been made  by using three types of diagnostics:\\
(1) The theoretical development of an \hii region has triggered the empirical classification from initially hyper-compact \hii (HCH\mbox{\sc ~ii}) regions to ultra-compact \hii (UCH\mbox{\sc ~ii}) regions, compact \hii regions, and then classical/developed \hii regions (see \cite{keto03} and references therein);\\
(2) Since the warm inner parts of high-mass protostellar envelopes evolve with time, the physical and chemical properties of a hot core (e.g., its size, temperature, molecular abundances, and associated
masers) can in principle be used as a clock (e.g. Helmich \& van Dishoeck \cite{HvD97}; Garay \& Lizano \cite{GL99});\\
(3) Inspired by the sequence from Class~0 to Class~I observed for low-mass protostars (e.g. Andr\'e et al.\cite{AWB00} and Sect.~3.1 of the chapter on low-mass star formation), the ratio of submillimeter to bolometric luminosity has also been employed for OB-type protostellar objects (e.g., Molinari et al. \cite{mol98a}; Motte et al. \cite{MSL03}).\\
Following these empirical diagnostics, objects associated with the first phase of high-mass star formation have been called massive starless or pre-stellar cores, massive cold molecular cores, or even infrared-dark clouds. In the subsequent phase, they have been named massive protostars,  high-mass protostellar objects, or hot molecular cores. The better-known final phase would correspond to \hii regions being from hyper-compact to classical. Unfortunately, the above evolutionary scenarios are not quantitative enough to constrain models of high-mass star formation. Most of the questions related to the earliest phases of this process (i.e. the two first phases mentioned above) remain: What are their lifetimes? How do their luminosity and spectral energy distributions (SEDs) evolve with time? ...

\subsubsection*{What are the initial conditions of OB star formation?}

Since the earliest phases of high-mass star formation have long escaped observations, the characteristics of the initial cloud core or cloud structure that will form a high-mass star remain unknown. Here are a few basic questions we need to answer: Do singular massive pre-stellar cores exist? Are these hotter, denser, more turbulent, more magnetized, or more dynamic than their low-mass analogs? Do their characteristics permit the development of accretion rates $\sim$100 times higher than those of lower-mass protostars? Will they allow an effective coalescence of protostars in their inner parts?\\
Besides, since high-mass star-forming regions are more exposed to shock waves (powered by, e.g., OB stellar winds and ionization fronts), the formation of high-mass stars may be more commonly triggered than that of low-mass stars. When the impact of external triggers on clouds is constructive (see Elmegreen \cite{elme98} for a review), how does the initial high-mass pre-stellar core form?  and do external disturbances precipitate its collapse?\\
 
In the following, we review the major studies that have searched for the precursors of \hii regions and describe the future \emph{Herschel} key programmes dedicated to high-mass star formation. For a meaningful comparison of the massive YSOs identified by these studies (see Table~1), we choose to use the terminology recommended by Williams et al. (\cite{WBMK00}): ``clumps'' are  $\sim$1~pc cloud structures, ``dense cores'' have  $\sim$0.1~pc and ``condensations''  $\sim$0.01~pc sizes. 

\begin{table*}[htbp]
\caption[]{Comparison of millimeter cloud structures of a few reference studies}
\centering
\begin{tabular}{|l|ccc|cc|}
\hline
& HMPOs	& IRDCs & Cygnus~X	& Low-mass	& $\rho$~Oph\\
& clumps	& clumps	& dense cores	& dense cores & condens.\\
\hline
\hline
Diameter (pc)	& 0.5		& 0.5		& 0.13	& 0.08 	& 0.007\\
Mass (\msun)		& 290	& 150	 & 91 		& 4.7 	& 0.15 \\
$n_{\mbox{\tiny H$_2$}}$ (cm$^{-3}$)	& $8\times 10^3$ 	& $6\times 10^3$ 	 & $2\times 10^5$ & $3\times 10^4$	& $2\times 10^6$ \\
$d_{\mbox{\scriptsize Sun}}$ (kpc) & 0.3-14  & 1.8-7.1 & 1.7 & 0.14-0.44 & 0.14\\
References  & (1) & (2) & (3) & (4), (5) & (5)\\
\hline
\end{tabular}
References: (1) Beuther et al. (\cite{beut02}); (2) Rathborne et al. (\cite{rath06}); (3) Motte et al. (\cite{mott07}); (4)  Ward-Thompson et al. (\cite{WMA99}); (5) Motte et al. (\cite{MAN98}).
\end{table*}

\subsection{Galaxy-wide surveys of high-luminosity YSOs}

Newborn massive stars that have developed an \hii region are strong free-free emitters at centimeter wavelengths and have thus been studied in great details for several decades (see the review by, e.g., Churchwell \cite{chur02}). In \cite{WC89}, Wood \& Churchwell started searching for the youngest \hii regions by using the Galaxy-wide survey of high-luminosity infrared sources provided by \emph{IRAS}. They applied the color-color criteria illustrated in  Fig.~1 to select red \emph{IRAS} sources that could correspond to young stellar objects with a stellar embryo more massive than $8~\msun$. The resulting catalog contains 1646 sources spread near and far across the Galaxy. Most of these sources are indeed UC\hii regions but some of them could even be in the earlier protostellar phase.

\begin{figure}[htbp]
\centerline{\includegraphics[width=10.2cm,angle=270]{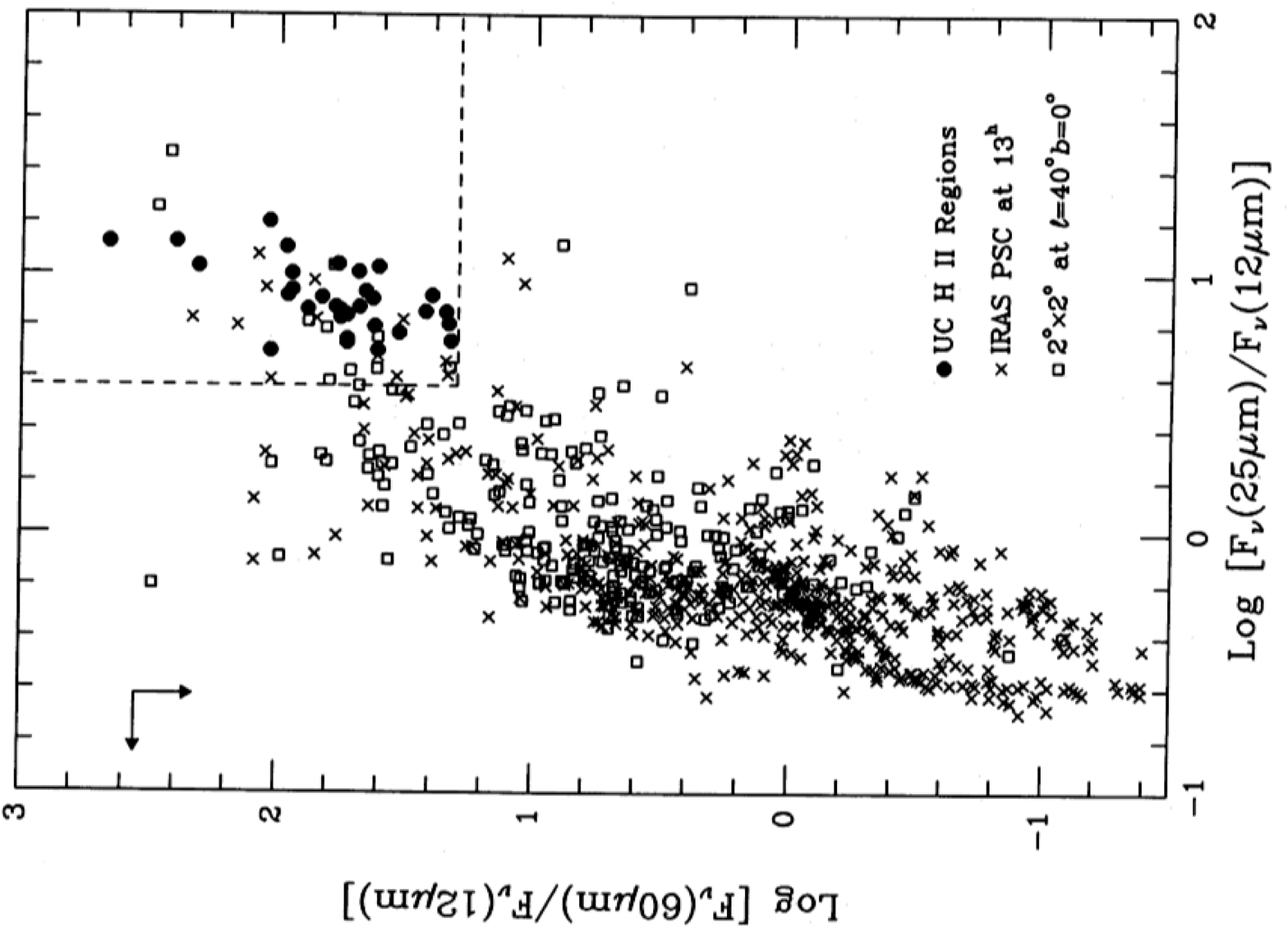}}
\caption{Taken from Wood \& Churchwell (\cite{WC89}): Color-color diagram for part of the \emph{IRAS} point sources catalog and a few well-known UC\hii regions. The upper-right box locates sources satisfying the color-color criteria of Wood \& Churchwell (\cite{WC89}): $Log(F_{60}/F_{12}) > 1.3$ and $Log(F_{25}/F_{12}) > 0.57$.}
\end{figure}

Several authors have searched for protostellar objects within the Wood \& Churchwell catalog of \emph{IRAS} sources (e.g. Bronfman et al. \cite{BNM96}; Plume et al. \cite{plum97}; Molinari et al. \cite{mol00}; Beuther et al. \cite{beut02}). They have investigated their association with dense gas through for example CS or millimeter continuum detection, with a hot core through detection of complex molecules or masers emission, and checked the absence of any \hii region via no or weak emission at centimeter wavelengths. The best-studied sample of Sridharan/Beuther contains 69 objects called HMPOs (for High-Mass Protostellar Objects) which are located at 300~pc to 14~kpc from the Sun. Following the terminology of Williams et al. (\cite{WBMK00}), the median HMPO (see Table~1) is a ``clump'', i.e. a $\sim$1~pc cloud structure hosting individual protostars and starless cores with expected 0.01--0.1~pc sizes. As usual with \emph{IRAS}-selected samples, the \emph{IRAS} sources are either closely or only loosely associated with the cloud clumps called here HMPOs. The former HMPOs are thus good candidates to contain high-luminosity infrared protostars while the nature of the latters remains to be determined.

As shown above, our understanding of high-mass star formation has been, for long, exclusively based on follow-up studies of bright sources found by \emph{IRAS}. However, if the high-mass star formation process goes through cold, low-luminosity phases reminiscent of those of low-mass pre-sellar cores and Class~0s (cf. Andr\'e et al. \cite{AWB00}), our knowledge has long been biased against these earliest phases. The two following sections summarize the recent discoveries made in this area.

\subsection{First discoveries of massive infrared-quiet YSOs}

The first sources which were identified as good candidates for being infrared-quiet precursors of high-mass stars have been found by two methods:
\begin{itemize}
\item The first one uses high-density tracers (often submillimeter continuum) to map the surroundings of massive YSOs associated with well-known \hii regions, H$_2$O or CH$_3$OH masers, or \emph{IRAS} sources. Many of these mappings have serendipitously revealed dense and massive cloud fragments which remain undetected at mid-infrared wavelengths (e.g. Motte et al. \cite{MSL03}; Hill et al. \cite{hill05}; Klein et al. \cite{klein05}; Sridharan et al. \cite{srid05}; Beltr\'an et al. \cite{belt06}; Thompson et al. \cite{thom06}). These studies are evidently plagued by very low-number statistics and large inhomogeneity since  the cloud fragments identified this way have sizes ranging from 0.1~pc to more than 1~pc.
\item A second method is to search for sources seen in absorption against the diffuse mid-infrared background of square degrees images taken by the \emph{ISO}, \emph{MSX}, and most recently \emph{Spitzer} satellites. Indeed, these absorption features, often called InfraRed Dark Clouds (IRDCs), could be the footprints of cold cloud structures as generally confirmed by maps of high-density cloud tracers (see, e.g., Fig.~2). These IRDCs surveys provide large samples of infrared-quiet sources generally located at large and inhomogeneous distances from the Sun (e.g. P\'erault et al. \cite{per96}; Egan et al. \cite{egan98}; Carey et al. \cite{car00}; Hennebelle et al. \cite{hen01}; Rathborne et al. \cite{rath06}).
Even in the most recent study by Rathborne et al. (\cite{rath06}), the selected sources (called in their paper IRDCs cores) have the size of clumps that could harbor several infrared-quiet protostars and/or pre-stellar cores (see Table~1). 
\end{itemize}

\begin{figure}[htbp]
\includegraphics[width=12.5cm]{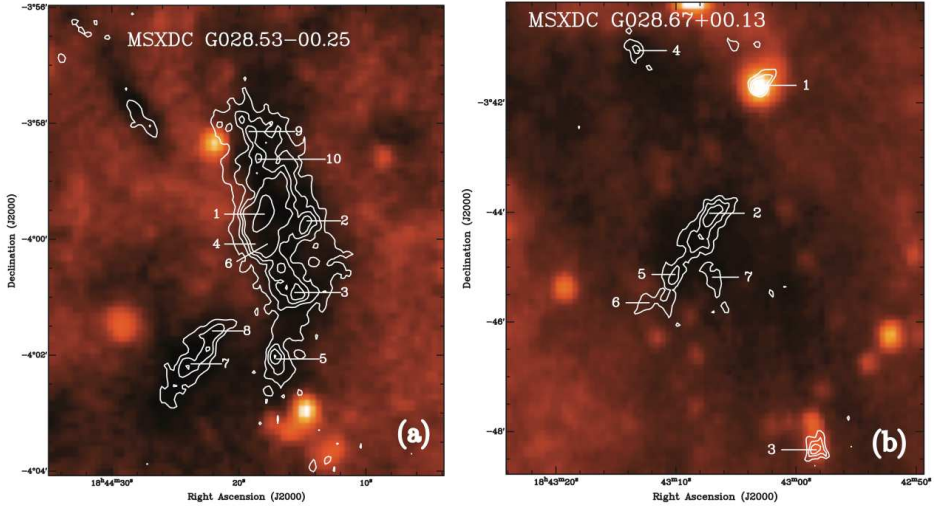}
\caption{Taken from Rathborne et al. (\cite{rath06}): Two InfraRed Dark Clouds detected in \emph{MSX} 8~$\mu$m images (color scale) as absorption features and confirmed by MAMBO-2 1.2~mm continuum emission (contours) as filamentary clouds splitting into several clumps.}
\end{figure}

The sources identified above are definitively colder than the high-luminosity infrared sources discussed in Sect.~3.2. They could be either starless clumps or infrared-quiet protostellar objects, depending on the existence of protostellar activity signatures such as outflows, hot cores, or masers. However, very few of these infrared-quiet YSOs have been surveyed for protostellar activity signatures. Moreover, given their large size and moderate mass (see Table~1), many of them are probably not dense enough to form high-mass stars in the near future. In summary, no high-mass pre-stellar core have yet been convincingly identified by the above methods and only a handful of sources have been studied with enough spatial resolution, SED coverage, and follow-up studies to qualify as high-mass equivalent of Class~0 protostars (Hunter et al. \cite{hunt98}; Molinari et al. \cite{mol98a}; Sandell \cite{sand00}; Garay et al. \cite{gara02}; Sandell \& Sievers \cite{SS04}).

\subsection{Imaging survey of entire complexes forming OB stars: unbiased census of both infrared-quiet and infrared-bright massive YSOs}

\begin{figure}[htbp]
\vskip -0.4cm
\includegraphics[height=13cm, width=10cm,angle=270]{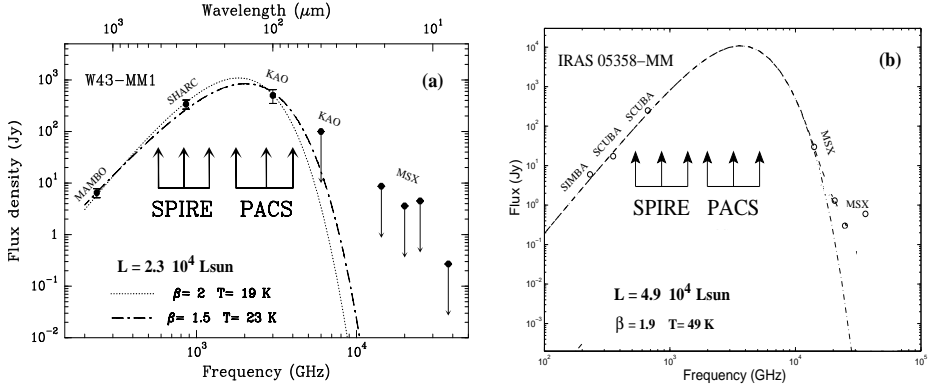}
\vskip -4cm
\caption[]{Spectral energy distributions of {\bf a)} W43-MM1, a clump which is harboring an infrared-quiet high-mass protostar (Motte et al. \cite{MSL03}) and {\bf b)} the high-luminosity infrared protostar IRAS~05358-MM (Minier et al. \cite{mini05}) compared with graybody models.}
\end{figure}

To make progress, one needs to search, in a systematic and unbiased way, for high-mass analogues of pre-stellar cores, Class~0 and Class~I protostars. If they exist, they should be small-scale ($0.01-0.1$~pc) cloud fragments particularly dense (volume-averaged density $n_{\mbox{\tiny H$_2$}}=10^5-10^7$~cm$^{-3}$) to permit the formation of one massive star. They thus should be best detected via far-infrared or (sub)millimeter dust continuum, which happens to be the wavelengths domain of \emph{Herschel} (see Fig.~3). To achieve sufficient spatial resolution and statistics, it is judicious to focus on the closest molecular cloud complexes which are actively forming OB stars. Actually, S. Bontemps (in prep.) recently identified eight such complexes by using CO surveys and a near-infrared extinction image of our Galaxy which was built from the stellar reddening measured by 2MASS. Since these complexes are located at intermediate distances (0.7 to 3~kpc), their study ensures reasonable spatial resolution with current (sub)millimeter facilities (\emph{HPBW}$=11''$ with MAMBO-2, $9''$ with SHARC-II, $18''$ with LABOCA, and $8''$ with ArT\'eMiS) and the future SPIRE and PACS imagers of \emph{Herschel} (\emph{HPBW}$=5''-36''$). Moreover, the amount of molecular gas contained in these cloud complexes should statistically allow studying the precursors of OB stars with masses up to $50~\msun$. Therefore, multi-tracer studies of such complexes are expected to provide, with $\sim$0.1~pc spatial resolution, more statistically significant and more homogeneous samples of massive YSOs than any of the studies mentioned above.
 
\begin{figure}
\includegraphics[angle=0,width=12.3cm]{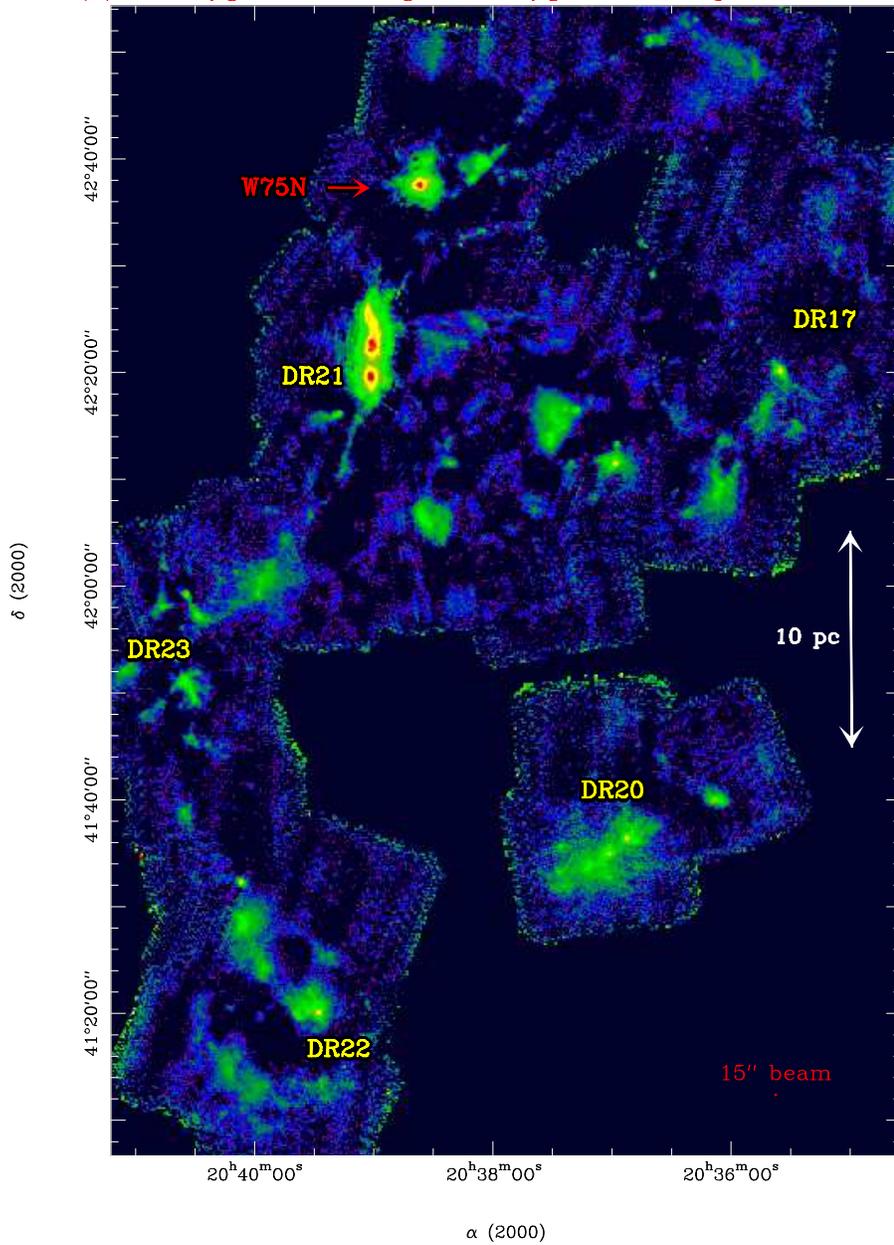}
\vskip -0.2cm
\caption[]{Taken from Motte et al. (\cite{mott07}): Millimeter continuum imaging of Cygnus~X North obtained with the MAMBO-2 camera installed at the IRAM 30~m telescope. This 1.2~mm image is sensitive to 0.09--5~pc cloud structures, i.e. from the scale of dense cores ($\sim$0.1~pc) to that of clumps ($\sim$1~pc).  This study has identified 17 new good candidates for hosting a massive infrared-quiet protostar.}
\end{figure}

Among the most active star-forming complexes at less than 3~kpc, Cygnus~X is the one that has recently caught most of the attention. According to Schneider et al. (\cite{schn06}), this massive ($4\times 10^6~\msun$) molecular complex is tightly associated with several OB associations (the largest being Cyg~OB2) and is located at only 1.7~kpc from the Sun. The high-density clouds of Cygnus~X have been completely imaged in millimeter continuum, CS, and N$_2$H$^+$ and then compared with mid-infrared images of \emph{MSX}. Numerous follow-up observations of the best candidate progenitors of high-mass stars have been performed to constrain their characteristics and evolutionary status. The millimeter imaging survey of the entire Cygnus~X molecular complex has revealed hundreds of massive dense cores, among which $\sim$42 probable precursors of high-mass stars (see Fig. 4;  Motte et al. \cite{mott07}). In terms of size and density, these dense cores are indeed intermediate cloud structures between ``clumps'' like IRDCs or HMPOs and ``condensations'' (see Table~1). The Cygnus~X dense cores have higher density than their low-mass counterparts (see Table~1), making them good candidate sites for high-mass star formation. The \emph{MSX} fluxes emitted by the Cygnus~X dense cores have been used to separate high-luminosity ($>10^3~\lsun$) infrared YSOs and massive infrared-quiet objects. Surprisingly, SiO outflow measurements have shown that all the infrared-quiet cores of Cygnus~X are harboring (at least) one massive infrared-quiet protostar. Finally, Motte et al. (\cite{mott07}) qualify 17 cores as good candidates for hosting massive infrared-quiet protostars, while up to 25 cores potentially host high-luminosity infrared protostars. Interestingly enough, no starless core was found within this first unbiased and homogeneous sample of high-mass YSOs.

Since the Motte et al. (\cite{mott07}) sample is derived from a single molecular complex and covers every embedded phase of  high-mass star formation, it gives the first statistical estimates of their lifetime (see Table~2). Estimated relatively to the known age and numbers of OB stars in Cyg~OB2, the lifetimes of high-mass protostars and pre-stellar cores in Cygnus~X are $\sim$3$\times 10^4$~yr and $< 10^3$~yr. In rough agreement with their free-fall time estimates (see Eq.~2.3 in the chapter on low-mass star formation), these statistical lifetimes are one and two order(s) of magnitude smaller, respectively, than what is found in nearby, low-mass star-forming regions (Kirk et al. \cite{KWA05}; Kenyon \& Hartmann \cite{KH95}). These results further suggest that high-mass pre-stellar and protostellar cores are in a highly dynamic state, as expected in a molecular cloud where turbulent processes dominate.

\begin{table*}[htbp]
\caption[]{Lifetime/age estimates (in years) of massive YSOs in Cygnus~X}
\centering
\begin{tabular}{|l|ccc|cc|}
\hline
&  Pre-stellar	& IR-quiet		& IR	& \hii 	& OB\\
&  cores		& protostars	& protostars		& regions	& stars\\
\hline
\hline
Statistical lifetime
	& $\le 8 \times 10^2$ 	& $1\times 10^4$ 	& $2 \times 10^4$ & $6\times 10^5$	& $2 \times 10^6$ \\
Free-fall time
	& $\sim 9\times 10^4$ 	& \multicolumn{2}{c|}{$8\times 10^4$} & & \\
Low-mass lifetimes
          & $2\times 10^5$	& $2\times 10^4$	& $2\times 10^5$ & & \\
\hline
\end{tabular}
\end{table*}

\subsection{\emph{Herschel} key programmes dedicated to the earliest phases of high-mass star formation}

As suggested by our review, studying the cold earliest stages of high-mass star formation is becoming a ``hot topic''. \emph{Herschel} is perfectly suited for such studies thanks to its broad wavelength coverage of cold SEDs (see Fig.~3), its unprecedented resolution at far-infrared wavelengths, its high scanning speed, and its spectral access to line diagnostics for \hii regions and photo-dissociation regions, as well as H$_2$O cooling lines. Three Guaranteed Time and one major Open Time Key Programmes are worth mentioning:
\begin{itemize}
\item ``HOBYS'' (the \emph{Herschel} imaging survey of OB Young Stellar objects) is a Guaranteed Time Key Programme jointly proposed by the SPIRE and PACS consortia, and the Herschel Science Centre (see http://starformation-herschel.iap.fr/hobys/). Coordinated by Motte, Zavagno, and Bontemps, it is the only \emph{Herschel} Key Programme which is exclusively dedicated to high-mass star formation. Its wide-field photometry part with both SPIRE and PACS aims at making the census of massive YSOs in the richest molecular complexes at less than 3~kpc. This survey will provide the statistical base to derive the lifetimes of massive YSOs (see Sect.~3.4 for Cygnus X). SPIRE-PACS unprecedented wavelength coverage will give robust measurements of the basic properties (bolometric luminosity and mass) of each YSO. This is crucial to build quantitative evolutionary diagrams such as the $M_\mathrm{env}-L_\mathrm{bol}$ diagram for lower mass young stellar objects (e.g. Andr\'e \& Montmerle \cite{AM94}). The ``HOBYS'' project is expected to multiply by more than 10 the number of high-mass analogs of Class~0 protostars known to date (see Sect.~3.3).\\
In complement, the detailed spectroscopic and photometric part of the ``HO\-BYS'' Key Programme will study with PACS clear examples of triggered star formation (Deharveng et al. \cite{DZC05}). The selected regions are hot photodissociation regions and will be observed with SPIRE in a companion Key Programme (cf. Abergel et al., using both spectroscopic and photometric modes). The combination of these observations will allow to characterize the second generation of massive YSOs and constrain the initial condition of star formation at the periphery of developped \hii regions. Such detailled studies on individual sources will allow us to better understand the positive feed-back of massive stars on their surrounding clouds. The larger view obtained with the SPIRE and PACS imaging surveys of the molecular complexes selected for ``HOBYS'' will allow us to assess the impact of external triggers and, in particular, the importance of the collect and collapse scenario (Elmegreen \& Lada \cite{EL77}).
\item ``WISH'' (Water in Star-forming regions with \emph{Herschel}) is a Key Programme dedicated to  water lines which is proposed by the HIFI consortium and coordinated by van Dishoeck. Water is a very important chemical specie which is only observable from space. It plays an important role in the cooling of the high-density interstellar medium and is expected to be one of the best tracers of the physical properties and kinematics in the innermost regions of the collapsing envelopes. Both low- and high-mass protostars will be observed with HIFI and PACS in the spectrometric mode. In particular, as many as 24 massive YSOs at different evolutionary stages will be extensively studied, for the first time, in water lines (high-mass star formation part, coordinated by van der Tak, Herpin, and Wyrowski).
\item ``The earliest phases of low- and high-mass star formation'' (by Henning et al.) is a Key Programme aimed at characterizing the physical conditions of selected low- and high-mass YSOs through small SPIRE and PACS maps. Among their goals, they plan to derive 3D models of pre-stellar cores and protostars, as well as survey the low-mass star population forming in the vicinity of high-mass protostars or UC\hii regions.
\newpage
\item ``Hi-GAL'' (the \emph{Herschel} infrared Galactic Plane Survey by Molinari et al.) is one of the largest open time key programmes since it proposes to image the whole Galactic plane with SPIRE and PACS. The recent near-infrared to millimeter Galactic surveys such as Spitzer-GLIMPSE (Benjamin et al. \cite{ben03}) or ATLAS-GAL (Schuller et al.) have demonstrated the richness of large-scale views of star formation throughout our Galaxy. If accepted, Hi-GAL will provide a complete survey of massive YSOs ranging from the scale of dense cores ($\sim$0.1~pc) to that of clouds ($\sim$10~pc) and located in various environments (solar neighborhood, Galactic center, molecular ring, interarm). Among other things, Hi-GAL is expected to give access to the global properties of star formation such as the star formation rate throughout the Galaxy or the statistical importance of external triggers.
\end{itemize}

\section{Conclusion:}

With the advent of the \emph{Herschel} satellite and soon after the ALMA interferometer we are entering a very promissing era for the studies of the earliest phases of star formation. Large-scale imaging surveys complemented by high-resolution studies will, without any doubt, revolutionize our knowledge of high-mass star formation.


\end{document}